\documentclass[reviewcopy]{elsarticle}

\usepackage[reviewcopy]{adndt}
\usepackage{longtable}

\usepackage{amsmath}

\usepackage{amssymb}

\biboptions{square,sort&compress}
\bibpunct[]{[}{]}{,}{n}{}{;}
\begin{document}

\begin{frontmatter}

\journal{Atomic Data and Nuclear Data Tables}

%% Author, fill in article title here

\title{Radiative rates for E1, E2, M1, and M2 transitions in  F-like ions with 37 $\le$ Z $\le$ 53}%Template for \normalfont\textsc{Atomic Data and Nuclear Data Tables}}

%% Fill in author list here
  \author[One]{Kanti M. Aggarwal\fnref{}\corref{cor1}}
  \ead{K.Aggarwal@qub.ac.uk}

  \author[One]{Francis P. Keenan}%\fnref{}}

 % \author[Two]{C. Author} 

  \cortext[cor1]{Corresponding author.}
%  \fntext[X]{{\em e-mail address}: K.Aggarwal@qub.ac.uk}
%  \fntext[Y]{{\em e-mail address}: F.Keenan@qub.ac.uk}

  \address[One]{Astrophysics Research Centre, School of Mathematics and Physics, Queen's University Belfast,\\Belfast BT7 1NN,
Northern Ireland, UK}

%  \address[Two]{Second Address First Line\\
%    Second Address Second Line}

\date{16.12.2002} %please do not use \today, use actual date of version

\begin{abstract}  
Calculations of energy levels, radiative rates and lifetimes are reported for 17 F-like ions with 37 $\le$ Z $\le$ 53. For brevity, results are only presented among the lowest 113 levels of the 2s$^2$2p$^5$, 2s2p$^6$, 2s$^2$2p$^4$3$\ell$, 2s2p$^5$3$\ell$, and 2p$^6$3$\ell$ configurations, although the calculations have been performed for up to 501 levels in each ion.   The general-purpose relativistic atomic structure package ({\sc grasp}) has been adopted for the calculations,   and radiative rates (along with oscillator strengths and line strengths) are listed  for all E1, E2, M1, and M2 transitions of the ions. Comparisons are made with  earlier available experimental and theoretical energies, although these are limited to only a few levels for most ions. Therefore for additional accuracy assessments, particularly for energy levels, analogous calculations have been performed with the Flexible Atomic Code ({\sc fac}), for up to 72~259 levels. Limited previous results are available for  radiative rates for comparison purposes, and no large discrepancy is observed for any transition and/or ion. 
\\ \\
{\em Received}: 5 November 2015, {\em Accepted}: 30 November  2015

\vspace{0.5 cm}
\end{abstract}

\end{frontmatter}

%%% Keywords and subject classification are not used in ADNDT 
%%%\begin{keywords}
%%%Insert list of keywords here.
%%%\end{keywords}

%%% The table of contents should start a new page. This command will
%%% automatically pull all the unstarred \section, \subsection and
%%% \subsubsection titles into the Contents. Starred versions need to be
%%% done manually using
%%%            \addcontentsline{toc}{[[sub]sub]section}{Section title}
%%% at the correct place. Examples are given below.

%%% The lists of data figures and data tables are created automatically
%%% by the \listofDfigures and \listofDtables commands.

\newpage

\tableofcontents
\listofDtables
\listofDfigures
\vskip5pc

%%%% Authors begin text of article here %%%

\section{Introduction}

Transitions of F-like  ions are prominent in high temperature  plasmas \cite{read1, read2} and are  useful for  diagnostics. While transitions of low Z elements are comparatively more important for the study of astrophysical plasmas, the heavier ones are of more interest in laboratory sources. Of particular interest are ions of the fifth row elements, because these are increasingly injected as impurities in tokamak fusion plasmas. Considerable attention has been paid to calculations of atomic data for low Z elements, such as energy levels and radiative decay rates -- see for example \cite{bh, cff1}. However, similar atomic data are (generally)  lacking for heavier ions,  although Sampson et al. \cite{sam} have performed calculations for a wide range of F-like ions with 22 $\le$ Z $\le$ 92 by using their Dirac-Fock-Slater (DFS) code. Their focus was on collisional data, but they also reported  oscillator strengths for electric dipole transitions, although  only from the lowest three levels of the 2s$^2$2p$^5$ and 2s2p$^6$ configurations  to the 110 excited levels of  2s$^2$2p$^4$3$\ell$, 2s2p$^5$3$\ell$ and 2p$^6$3$\ell$. Furthermore, they did not report energy levels, although these can be inferred from their tabulations of collision strengths, but only for a few ions, as for brevity they did not report data for all. Similarly, J\"{o}nsson et al. \cite{jag} have calculated energies and radiative rates (A-values) for a wide range of ions with 14 $\le$ Z $\le$ 74. They  adopted the general-purpose relativistic atomic structure package ({\sc grasp}) code \cite{grasp2k} and have included very large {\em configuration interaction} (CI) for the calculations. However, their results are restricted to the lowest 3 levels of the 2s$^2$2p$^5$ and 2s2p$^6$ configurations. For the same transitions, measurements for wavelengths have also been made in laser-produced plasmas for many F-like ions, i.e. 38 $\le$ Z $\le$ 50 \cite{read1, read2, hut, uf1, uf2}. Additionally, Zigler et al. \cite{zig} have identified 11  lines in the 6--7 $\rm\AA$ range of the (2s$^2$2p$^5$)~$^2$P -- (2s$^2$2p$^4$)~3s, 3d transitions of Rb~XXIX. However, for modelling applications a larger set of data are (preferably) required. With this in mind and the requirements for the developing ITER project, we have already reported energy levels and A-values for two ions, namely Kr~XXVIII \cite{kr} and Xe~XLVI \cite{xe}, and here we calculate similar data for all F-like ions with 37 $\le$ Z $\le$ 53.

As in our earlier work \cite{kr, xe}, we adopt the {\sc grasp} code for our calculations, and  for the optimisation of the orbitals  use the option of  `extended average level' (EAL),  in which a weighted (proportional to 2$j$+1) trace of the Hamiltonian matrix is minimised. However, our version of the code is slightly different from the one used by J\"{o}nsson et al. \cite{jag}, although all versions originate from the same source \cite{grasp0} and provide similar  results. This version has been revised by one of the authors (P. H. Norrington), is referred to as GRASP0 and is freely available at  {\tt http://web.am.qub.ac.uk/DARC/}. As in our earlier work for Kr and Xe ions, we report energies for the lowest 113 levels of the 2s$^2$2p$^5$, 2s2p$^6$, 2s$^2$2p$^4$3$\ell$, 2s2p$^5$3$\ell$, and 2p$^6$3$\ell$ configurations, although calculations have been performed for a much larger number of levels -- see section 2. Similarly, we list A-values for all transitions among these levels for four types, namely electric dipole (E1),  magnetic dipole (M1), electric quadrupole (E2), and magnetic quadrupole (M2). These results are required for the further calculation of lifetimes.

\section{Energy levels}

In our earlier work \cite{kr, xe}, CI was included among the basic 11 configurations, i.e 2s$^2$2p$^5$, 2s2p$^6$, 2s$^2$2p$^4$3$\ell$, 2s2p$^5$3$\ell$, and 2p$^6$3$\ell$, which generate a total of  113 levels. However, in the present paper we include an additional 27 configurations, which are:  2s$^2$2p$^4$4$\ell$, 2s2p$^5$4$\ell$, 2p$^6$4$\ell$, 2s$^2$2p$^4$5$\ell$, 2s2p$^5$5$\ell$, and 2p$^6$5$\ell$. These 38 configurations generate 501 levels in total and provide slightly more accurate results than those obtained with the basic 11. Energies for the lowest 113 levels, obtained with the inclusion of Breit and quantum electrodynamic effects,  are listed in Tables 1--17 for ions with 37 $\le$ Z $\le$ 53. We note that energies of the higher levels from other configurations lie {\em above} the listed 113 and there is no mixing for any ion considered here. 

For any calculation it is very important to assess accuracy so that results can be confidently applied in plasma modelling \cite{fst}. It is generally easier to assess the accuracy of energy levels if corresponding experimental data are available. Unfortunately, that is not the case for a majority of the levels for the ions  studied here, although a few measurements are available, such as by \cite{read1, read2, hut, uf1, uf2, zig} as already noted. The available experimental data have been assessed by the National Institute of Standards and Technology (NIST) personnel and their compiled results are available at the  website {\tt http://www.nist.gov/pml/data/asd.cfm}. 
 
Another method  of assessing  accuracy is to perform calculations by a different and independent approach. For this  we have adopted the {\em Flexible Atomic Code} ({\sc fac})  of  Gu  \cite{fac}. This is also a fully relativistic code and is available from the website {\tt https://www-amdis.iaea.org/FAC/}. The advantages of FAC are that it is  highly efficient to run and generally yields results comparable to those obtained  with other atomic structure codes, as has already been demonstrated in several of our earlier papers, including those on F-like Kr~XXVIII \cite{kr} and Xe~XLVI \cite{xe}. Therefore, the results obtained (FAC1) with the same CI as with {\sc grasp} are also listed in Tables 1--17. Both sets of energies, for all ions, agree very well (within 0.05 Ryd) and there is no serious discrepancy in level orderings. However, this result is fully expected and therefore  not  very useful for assessing the accuracy. For some ions, the inclusion of a very large CI  helps to  improve the accuracy of the energy levels, as already noted in our earlier work \cite{kr, xe}. We have hence performed another calculation (FAC2) with 38~089 levels. The levels additional to those of FAC1 arise from all possible combinations of the (2*5) 3*2, 4*2, 5*2, 3*1 4*1, 3*1 5*1, and 4*1 5*1 configurations. These results are also listed in Tables 1--17, but we discuss in detail, as an example, level energies for only Sr~XXX.

In comparison to FAC1, the energies from the FAC2 calculations are lower by up to 0.17 Ryd  ($\le$ 0.1\%) for many levels, but there is no (major) discrepancy in level orderings. More importantly, the FAC2 energies are consistently {\em lower} for all levels and therefore are assessed to be comparatively more accurate. Since the effect of the additional CI included in FAC2 is noticeable, this may increase with yet more CI. Therefore, we have performed another calculation (FAC3) by almost doubling the number of levels, specifically by including  72~259 levels in total, the additional ones arising from the (2*6) 6*1, 7*1, 8*1 and (2*5 3*1) 6*1,  7*1, and  8*1 configurations. The energies obtained for the lowest 113 levels of Sr~XXX from these FAC3 calculations are also listed in Table~2. However, the differences between the FAC2 and FAC3 energies are insignificant ($\le$ 0.013 Ryd). For many levels the FAC3 energies are lower than those from FAC2 but for a few the reverse is true -- see for example,   6, 14 and 18. Therefore, we can confidently state that energies for the lowest 113 levels of Sr~XXX (and other ions) have {\em converged}\, and the CI included in the FAC2 calculations is sufficient to obtain accurate results.

A major problem for most atomic structure calculations is the identification of level designations. Although levels of higher configurations do not mix with the lowest 113 of the F-like ions considered here, there is considerable mixing among themselves for a few. As an example, we list in Table~A all the levels of  Sr~XXX which are highly mixed. For this reason the level designations provided in Tables 1--17 should not be taken as definitive, as a few  may (inter)change depending on the  calculation with different codes and/or CI. We have attempted to identify the levels based on the strength of their eigenvectors, but for a few ambiguity remains, such as  6/38, 8/26 and 24/31. This is because in such cases a single eigenvector of a configuration state function (CSF) dominates in several levels.

As already stated, experimental energies for some levels are available on the NIST website for a few ions, namely Rb~XXIX, Sr~XXX and Mo~XXXIV. Energy levels for Sr~XXX have already been compared with theoretical results \cite{ajm}, which are similar to our calculations. Therefore, in Tables B and C we compare our results with those of NIST  for the common levels of Rb~XXIX and Mo~XXXIV, respectively. The NIST energy levels for Rb~XXIX have been compiled by Sansonetti \cite{san} and generally agree with our calculations, although differences for a few are up to  0.4 Ryd -- see for example levels 36, 51, 53, and 66  in Table~B. There is no trend, because for some our energies are higher and for a few lower. Additionally, there are some differences in the level designations and for this reason we have included mixing coefficients for these levels {\em plus} a few other which will help to explain the reason. As an example, level 40 is designated as 2s$^2$2p$^4$($^3$P)~$^4$F$_{5/2}$ (45 in Table~B) by NIST but is 2s$^2$2p$^4$($^1$S)3d ~$^2$D$ _{5/2}$ in our work.  Level 40 is well mixed with 45 and hence their labels can be interchanged, whereas 45 has a clear dominance of $\sim$48\%. Similar differences in the designations are found for  levels 38/46, 39/66 and 43/66.

Energies for a few levels for Mo~XXXIV have been compiled by Sugar and Musgrove \cite{sug} and are included in Table~C along with our corresponding results from GRASP. The differences for a few levels are up to 0.45 Ryd (0.2\%), particularly for level 43, i.e. 2s$^2$2p$^4$3d~$^4$F$_{3/2}$. Additionally, as for the levels of Rb~XXIX in Table~B, for Mo~XXXIV there are some differences in label designations between our results and NIST. The change of labels are for 30/46, 31/72, 43/72, and 45/46, and for this reason we have provided mixing coefficients for these levels in Table~C. Apart from these, many more levels are highly mixed and therefore there are always differences in the level designations between any two independent studies.

Finally, as  stated earlier,  J\"{o}nsson et al. \cite{jag} have reported energies for the 2s$^2$2p$^5$~$^2$P$^o_{1/2}$ and 2s2p$^6$~$^2$S$_{1/2}$ levels of many ions, and in Table~D we compare their plus available experimental results with our calculations with both {\sc grasp} and {\sc fac}. It is clear that, in general, the theoretical energies of J\"{o}nsson et al. \cite{jag} are closer to the  experimental ones, whereas our calculated energies with both codes are consistently lower by $\sim$ 0.02 Ryd  for the level 2s$^2$2p$^5$~$^2$P$^o_{1/2}$ and  $\sim$0.2 Ryd higher for 2s2p$^6$~$^2$S$_{1/2}$. This is because  J\"{o}nsson et al. have performed very large calculations by including up to 73~000 and 15~000 CSFs for the respective levels, whereas our calculations are comparatively modest keeping in mind the larger number of levels considered in the work. However, it is interesting to note that for a few ions our calculated energies are slightly closer to the measurements  -- see for example Z = 39 and 41 for 2s$^2$2p$^5$~$^2$P$^o_{1/2}$ and Z = 50 for  2s2p$^6$~$^2$S$_{1/2}$. To conclude, based on the comparisons shown in Tables~B, C and D we may confidently state that our energy levels listed in Tables 1--17 are accurate to $\sim$0.5\%.
 
\section{Radiative rates}\label{sec.eqs} 

In Tables 18--34 we present our calculated results  with the {\sc grasp} code for energies (wavelengths, $\lambda_{ji}$ in ${\rm \AA}$), radiative rates (A-values, in s$^{-1}$), oscillator strengths (f-values, dimensionless), and line strengths (S- values, in atomic units, 1 a.u. = 6.460$\times$10$^{-36}$ cm$^2$ esu$^2$)  for E1 transitions in F-like ions with 37 $\le$ Z $\le$ 53. However,  for E2, M1 and M2 transitions only the A-values are listed, because the corresponding results for f- or S-values can be  obtained using Eqs. (1-5) given in \cite{kr}. Additionally, we  list the ratio (R) of the  velocity (Coulomb gauge) and length (Babushkin gauge) forms which generally give an indication of the accuracy. The {\em indices} used to represent the lower and upper levels of a transition are defined in Tables 1--17. Furthermore,  for brevity only transitions from the lowest 3 to higher excited levels are listed in Tables 18--34, but  full tables are available online in the electronic version.

J\"{o}nsson et al. \cite{jag} have reported A-values for transitions among the lowest 3 levels of F-like ions, and in Table~E we compare our results for the 1--3 E1, 2--3 E1, 1--2 M1, and 1--2 E2 transitions. There is no discrepancy for any transition and/or ion and all results agree to better than 5\%. This is highly satisfactory and to a certain extent confirms the accuracy of our results. However, this comparison is very limited and therefore in Table~F we compare our f-values for three ions, namely Mo~XXXIV, Pd~XXXVIII and Sn~XLII with the earlier work of Sampson et al. \cite{sam},  for all E1 transitions from the ground level. Since the level orderings are not the same for all ions, the upper level J is also listed in this table which corresponds to Tables~6, 10 and 14 for the respective ions. For most transitions with significant f-values  (f $\ge$ 0.01) there is no discrepancy and both calculations agree within $\sim$10\%. However, for some weaker transitions the differences between the two sets of f-values are sizeable -- see for example, 1--97/98 of Mo~XXXIV, 1--90/97 of Pd~XXXVIII and 1--73/97 of Sn~XLII. This is because the weaker transitions are more susceptible to change with differing amount of CI and/or methods.

Finally, we discuss the ratio (R) of the  velocity and length  forms of the f-value to further assess  the  accuracy. For brevity, we only discuss results for transitions in Sr~XXX, although the same conclusions apply to all ions considered here. Among the lowest 113 levels, there are 2191 possible E1 transitions,  of which 385 have f $\ge$ 0.01, i.e. they are comparatively strong. For most of these 385 transitions, R is within 20\% of unity, although for about 25\% (104), R is up to a factor of 2 (such as 96--106, f = 0.020 and 105--113, f = 0.026), and only for one (58--106; f = 0.011) is R = 3. Furthermore, all such transitions have f $<$ 0.1. However, for a few very  weak transitions, R is up to several orders of magnitude, and examples include 5--111 (f $\sim$ 3$\times$10$^{-7}$), 14--43  (f $\sim$ 4$\times$10$^{-5}$) and 25--26  (f $\sim$ 6$\times$10$^{-9}$). As already stated, f-values for such weak transitions are highly variable with differing amount of CI, because of the cancellation and/or additive effects of different matrix elements, and hence are comparatively less reliable. However, due  to their small magnitudes, their contribution to the  modelling of plasmas is normally expected to be small.   Overall, based on this and other comparisons already discussed, our assessment of accuracy for the f- (and A-) values for a majority of strong transitions is $\sim$20\%, for all ions. 

\section{Lifetimes}

Once the A-values are known, the lifetime $\tau$ of a level $j$ can be easily determined as  1.0/$\Sigma_{i}$A$_{ji}$, where the summation includes results from all types of transitions, i.e. E1, E2, M1, and M2. Since this is a measurable quantity it helps to assess the accuracy of radiative rates, particularly when a single A-value for any type  of transition dominates. To our knowledge no measurements of $\tau$ are available for the levels of the F-like ions considered here, but in Tables 1--17  we have  listed our calculated results for future comparisons and  assessment of accuracy.  Previous theoretical results are available for only one ion, namely Sr~XXX \cite{ajm}, for which there is discrepancy for only level 2, i.e. 2s$^2$2p$^5$~$^2$P$^o_{1/2}$, and the $\tau$ of \cite{ajm} is underestimated by a factor of 2 as already explained \cite{kmq}.

\section{Conclusions}

Energies for the lowest  113 levels of the 2s$^2$2p$^5$, 2s2p$^6$, 2s$^2$2p$^4$3$\ell$, 2s2p$^5$3$\ell$, and 2p$^6$3$\ell$ configurations of  17 F-like ions with 37 $\le$ Z $\le$ 53 are reported. For the calculations the {\sc grasp} code has been adopted, although for an accuracy assessment the {\sc fac} is also employed with much more extensive CI. Based on several calculations with both codes and comparisons with available theoretical and experimental data, our energy levels are assessed to be accurate to better than 0.5\%, for all ions. However, for a few levels of each ion there is some ambiguity in level designations, because of very strong mixing with often one eigenvector of a CSF dominating in magnitude for several levels.

Radiative rates for E1, E2, M1, and M2 transitions are also reported among the lowest 113 levels, and our data are significantly more extensive than currently  available in the literature.  However, there is no major discrepancy for any transition and/or ion.   Based on several comparisons with different calculations, the accuracy of our A-values is assessed to be  $\sim$20\%, particularly for  strong transitions with large f-values, although for very weak transitions the reported A-values may be comparatively less reliable.   

For future comparisons, lifetimes for these levels are also listed although no measurements are currently available in the literature. However, previous theoretical values are available for the levels of Sr~XXX, and there is no discrepancy with our work, except for one level. Finally, calculations for energies have been made for up to  38~089 levels and for A-values among 501 levels, for all ions. Therefore, a larger set of data than presented here for any ion may be obtained on request from the first author (K.Aggarwal@qub.ac.uk).

%You are welcome to use BiBTeX with the \adndtbst\ bibliography
%style distributed with \adndtstyle\ package. This style comes very close
%to the journal style. Be sure to provide your
%\texttt{.bbl}  file (not the \texttt{.bib} file) with your submission.

%Please see \adndtguide\ for more instructions.

\ack
KMA  is thankful to  AWE Aldermaston for financial support. 

\begin{appendix}

\def\thesection{} % To get the appendix heading correct

\section{Appendix A. Supplementary data}% The Appendix itself

Owing to space limitations, only parts of Tables 18--34  are presented here, but full tables are being made available as supplemental material in conjunction with the electronic publication of this work. Supplementary data associated with this article can be found, in the online version, at doi:nn.nnnn/j.adt.2016.nn.nnn.

\end{appendix}

%%  All sections inside the appendix environment will be appendixes
%%  Subsections function normally in appendixes.

%\end{document}

\clearpage
\newpage

%The tables that are part of the introductory material should be located after the figures, one table per page.

\renewcommand{\baselinestretch}{1.0}
\footnotesize
\begin{longtable}{@{\extracolsep\fill}rllr@{}}
\caption{Mixing coefficients (MC) for some levels of Sr~XXX. Numbers outside and inside a bracket correspond to MC and the level, \\respectively. See Table~2 for the definition of all levels.}
Index  & Configuration           & Level              &  Mixing coefficients     \\  \\
 \hline\\
\endfirsthead\\
\caption[]{(continued)}
Index  & Configuration           & Level              &  Mixing coefficients   \\  \\
\hline\\
\endhead
   6 & 2s$^2$2p$^4$3s	      & $^2$S$  $$_{1/2}$  &   0.60( 38)-0.36( 13)+0.71(  6)			 \\
   8 & 2s$^2$2p$^4$($^3$P)3p  & $^2$D$^o$$_{5/2}$  &   0.56( 26)-0.47( 11)+0.49(  8)+0.40( 22)-0.25( 37) \\
  14 & 2s$^2$2p$^4$3p	      & $^4$D$^o$$_{1/2}$  &   0.54( 14)+0.55( 29)+0.29( 41)+0.57( 49)  	 \\
  18 & 2s$^2$2p$^4$($^3$P)3p  & $^2$P$^o$$_{3/2}$  &   0.55( 20)+0.38(  7)-0.25( 18)-0.21( 35)+0.65( 56) \\
  21 & 2s$^2$2p$^4$3d	      & $^4$D$  $$_{3/2}$  &  -0.64( 21)+0.46( 33)-0.27( 66)+0.20( 47)+0.40( 58)-0.25( 53)		 \\
  24 & 2s$^2$2p$^4$3d	      & $^4$P$  $$_{1/2}$  &   0.47( 42)-0.56( 24)-0.45( 31)+0.50( 59)  	 \\
  26 & 2s$^2$2p$^4$3p	      & $^4$D$^o$$_{5/2}$  &   0.76( 26)+0.50( 11)-0.38(  8)			 \\
  31 & 2s$^2$2p$^4$($^3$P)3d  & $^2$P$  $$_{1/2}$  &   0.63( 24)-0.57( 31)+0.47( 51)			 \\
  38 & 2s$^2$2p$^4$3s	      & $^4$P$  $$_{1/2}$  &  -0.57( 38)+0.44( 13)+0.68(  6)			 \\
  39 & 2s$^2$2p$^4$($^1$S)3d  & $^2$D$  $$_{3/2}$  &   0.56( 43)+0.47( 47)+0.21( 58)+0.59( 39)  	 \\
  40 & 2s$^2$2p$^4$3d	      & $^4$P$  $$_{5/2}$  &   0.50( 45)+0.34( 23)+0.22( 40)-0.27( 36)+0.67( 65) \\
  43 & 2s$^2$2p$^4$3d	      & $^4$F$  $$_{3/2}$  &  -0.45( 43)+0.63( 21)+0.38( 33)-0.25( 66)+0.29( 47)+0.26( 58)		 \\
  49 & 2s$^2$2p$^4$($^1$S)3p  & $^2$P$^o$$_{1/2}$  &  -0.44( 14)-0.52( 10)-0.22( 29)+0.65( 49)  	 \\
  56 & 2s$^2$2p$^4$($^1$S)3p  & $^2$P$^o$$_{3/2}$  &  -0.37( 20)-0.34(  7)-0.26( 17)+0.40( 30)+0.68( 56) \\
  64 & 2s2p$^5$($^3$P)3p      & $^2$D$  $$_{5/2}$  &   0.67( 74)-0.42( 69)+0.61( 64)			 \\
  65 & 2s$^2$2p$^4$($^1$S)3d  & $^2$D$  $$_{5/2}$  &  -0.33( 45)-0.33( 23)-0.33( 40)+0.39( 46)+0.69( 65) \\
  66 & 2s$^2$2p$^4$($^3$P)3d  & $^2$D$  $$_{3/2}$  &  -0.44( 43)-0.23( 21)-0.40( 66)-0.33( 47)+0.66( 39) \\
  68 & 2s2p$^5$3p    	      & $^4$D$  $$_{3/2}$  &  -0.45( 68)+0.50( 63)-0.37( 75)+0.61( 72)  	 \\
  71 & 2s2p$^5$3p    	      & $^4$P$  $$_{1/2}$  &   0.51( 80)-0.67( 71)-0.23( 73)+0.39( 96)-0.28( 98) \\
  72 & 2s2p$^5$($^3$P)3p      & $^2$P$  $$_{3/2}$  &  -0.54( 68)-0.66( 72)+0.46( 89)			 \\
  74 & 2s2p$^5$3p    	      & $^4$D$  $$_{5/2}$  &  -0.61( 74)-0.46( 69)+0.37( 64)+0.52( 97)  	 \\
  77 & 2s2p$^5$($^3$P)3p      & $^2$S$  $$_{1/2}$  &   0.51( 80)+0.62( 71)-0.37( 77)+0.36( 96)+0.23( 98) \\
  83 & 2s2p$^5$3d    	      & $^4$D$^o$$_{5/2}$  &  -0.41( 92)+0.71( 83)-0.56(101)			 \\
  85 & 2s2p$^5$($^3$P)3d      & $^2$D$^o$$_{5/2}$  &  -0.45( 92)+0.46(101)-0.38( 95)+0.65( 85)  	 \\
  88 & 2s2p$^5$($^3$P)3d      & $^2$D$^o$$_{3/2}$  &   0.46(102)-0.36( 93)-0.28( 81)-0.63( 88)+0.36(100)+0.21(108)		 \\
  90 & 2s2p$^5$3d    	      & $^4$D$^o$$_{1/2}$  &   0.68( 90)-0.71( 91)				 \\
  91 & 2s2p$^5$($^3$P)3d      & $^2$P$^o$$_{1/2}$  &   0.54( 90)+0.60( 91)+0.58(106)			 \\
  92 & 2s2p$^5$3d    	      & $^4$F$^o$$_{5/2}$  &  -0.51( 92)-0.24( 83)-0.24( 95)-0.59( 85)+0.52(103) \\
  93 & 2s2p$^5$3d    	      & $^4$D$^o$$_{3/2}$  &   0.33(102)-0.44( 93)-0.26( 81)+0.46( 88)-0.40(100)+0.29(108)-0.43(105)	 \\
  94 & 2s2p$^5$3d    	      & $^4$D$^o$$_{7/2}$  &  -0.50( 84)-0.51( 94)+0.47( 87)+0.52(104)  	 \\
 101 & 2s2p$^5$3d    	      & $^4$P$^o$$_{5/2}$  &  -0.44( 92)-0.54( 83)-0.41(101)+0.49( 95)+0.28( 85) \\
 104 & 2s2p$^5$($^1$P)3d      & $^2$F$^o$$_{7/2}$  &   0.36( 84)+0.31( 94)-0.21( 87)+0.85(104)  	 \\
  \\ \hline  											       								   					      
\end{longtable}

%\vspace*{0.5 cm}                                                                               
                               
\begin{flushleft}                                                                               
                               
{\small
                                                                               
}                                                                                               
                               
\end{flushleft} 
%}                        
%\end{document}

\clearpage
\newpage
\renewcommand{\baselinestretch}{1.0}
\footnotesize
\begin{longtable}{@{\extracolsep\fill}rllrrrr@{}}
\caption{Comparison of some energy levels of Rb~XXIX (in Ryd) and their mixing coefficients (MC).  Numbers outside and inside a bracket correspond to MC and the level, respectively. See Table 1 for definition of all levels.}
Index  & Configuration             & Level              &  NIST       &   GRASP     &  Mixing coefficients  \\  \\
 \hline\\
\endfirsthead\\
\caption[]{(continued)}
Index  & Configuration             & Level              &  NIST       &   GRASP     &  Mixing coefficients  \\  \\
\hline\\
\endhead 
    1  &  2s$^2$2p$^5$  	   &   $^2$P$^o_{3/2}$   &    0.000   &     0.0000  &	1.00(  1)	       \\
    2  &  2s$^2$2p$^5$  	   &   $^2$P$^o_{1/2}$   &    4.596   &     4.5872  &	1.00(  2)	       \\
    3  &  2s2p$^6$		   &   $^2$S$  _{1/2}$   &   18.325   &    18.4716  &	1.00(  3)	       \\
    5  &  2s$^2$2p$^4$3s	   &   $^2$P$  _{3/2}$   &  135.186   &   135.0347  &	0.40(  9)+0.77(  5)+0.49( 16)				       \\
   15  &  2s$^2$2p$^4$3s	   &   $^2$D$  _{5/2}$   &  140.153   &   140.0749  &	0.46(  4)-0.89( 15)					       \\
   36  &  2s$^2$2p$^4$3d	   &   $^4$P$  _{3/2}$   &  145.775   &   145.3680  &  -0.58( 36)-0.52( 66)+0.27( 57)+0.35( 54)-0.37( 39)	       \\
   38  &  2s$^2$2p$^4$($^3$P)3d    &   $^2$D$  _{5/2}$   &	      &   145.6587  &	0.48( 64)-0.47( 46)+0.53( 38)+0.35( 56)+0.24( 53)+0.21( 40)    \\
   39  &  2s$^2$2p$^4$($^1$S)3d    &   $^2$D$  _{3/2}$   &	      &   146.1513  &	0.57( 43)+0.46( 47)+0.20( 57)+0.59( 39) 		       \\
   40  &  2s$^2$2p$^4$($^1$S)3d    &   $^2$D$  _{5/2}$   &  146.249   &   146.4424  &	0.51( 45)+0.34( 24)+0.23( 64)-0.26( 38)+0.65( 40)	       \\
   43  &  2s$^2$2p$^4$3d	   &   $^4$F$  _{3/2}$   &  148.855   &   148.7185  &  -0.44( 43)+0.63( 23)+0.39( 36)-0.26( 66)+0.29( 47)+0.27( 57)    \\
   45  &  2s$^2$2p$^4$3d	   &   $^4$F$  _{5/2}$   &	      &   149.1016  &	0.69( 45)-0.38( 64)+0.49( 38)+0.24( 56)-0.21( 53)	       \\
   46  &  2s$^2$2p$^4$($^3$P)3d    &   $^2$F$  _{5/2}$   &  149.220   &   149.2505  &	0.26( 24)+0.52( 64)+0.76( 46)+0.22( 38) 		       \\
   51  &  2s$^2$2p$^4$($^1$D)3d    &   $^2$S$  _{1/2}$   &  150.323   &   150.0290  &  -0.47( 27)-0.35( 59)+0.81( 51)				       \\
   53  &  2s$^2$2p$^4$($^1$D)3d    &   $^2$D$  _{5/2}$   &  150.496   &   150.0662  &	0.27( 64)+0.24( 38)-0.45( 56)-0.77( 53) 		       \\
   54  &  2s$^2$2p$^4$($^1$D)3d    &   $^2$P$  _{3/2}$   &  150.396   &   150.2640  &  -0.28( 43)+0.36( 36)+0.32( 47)-0.22( 57)+0.76( 54)	       \\
   57  &  2s$^2$2p$^4$($^1$D)3d    &   $^2$D$  _{3/2}$   &  150.769   &   151.0617  &	0.22( 43)-0.54( 66)-0.75( 57)				       \\
   66  &  2s$^2$2p$^4$($^3$P)3d    &   $^2$D$  _{3/2}$   &  155.690   &   155.2674  &  -0.43( 43)-0.23( 23)-0.40( 66)-0.33( 47)+0.67( 39)	       \\
\\  \hline            								                	 
\end{longtable}   								   					       
			      							   					       
%\vspace*{0.5 cm}													       

%}													       
														       
\begin{flushleft}													       
{\small
NIST:  {\tt http://www.nist.gov/pml/data/asd.cfm} \\
GRASP: Present results with the {\sc grasp} code with 501 level calculations  \\	       
}															       
\end{flushleft} 

%\end{document}

\renewcommand{\baselinestretch}{1.0}
\footnotesize
\begin{longtable}{@{\extracolsep\fill}rllrrrr@{}}
\caption{Comparison of some energy levels of Mo~XXXIV (in Ryd) and their mixing coefficients (MC).  Numbers outside and inside a bracket correspond to MC and the level, respectively. See Table 6 for definition of all levels.}
Index  & Configuration             & Level              &  NIST       &   GRASP     &  Mixing coefficients  \\  \\
 \hline\\
\endfirsthead\\
\caption[]{(continued)}
Index  & Configuration             & Level              &  NIST       &   GRASP     &  Mixing coefficients  \\  \\
\hline\\
\endhead 
    1 & 2s$^2$2p$^5$		&  $^2$P$^o_{3/2}$  &	 0.0000      &   0.0000 &   1.00(  1)  		  \\
    2 & 2s$^2$2p$^5$		&  $^2$P$^o_{1/2}$  &  8.0766 &   8.0624 &   1.00(  2)  		  \\
    3 & 2s2p$^6$		&  $^2$S$  _{1/2}$  & 24.1969 &  24.3494 &   1.00(  3)  		  \\
   30 & 2s$^2$2p$^4$($^3$P)3d	&  $^2$D$  _{5/2}$  &	      & 194.6613 &   0.47( 71)-0.45( 46)+0.53( 30)+0.36( 53)+0.30( 51)+0.22( 35)  \\
   31 & 2s$^2$2p$^4$($^1$S)3d	&  $^2$D$  _{3/2}$  &	      & 195.1778 &   0.50( 43)+0.49( 47)+0.24( 58)+0.58( 31)			  \\
   43 & 2s$^2$2p$^4$3d  	&  $^4$F$  _{3/2}$  &200.2783 & 200.7266 &  -0.49( 43)+0.62( 19)+0.36( 28)-0.21( 72)+0.31( 47)+0.25( 58)  \\
   45 & 2s$^2$2p$^4$3d  	&  $^4$F$  _{5/2}$  &200.8979 & 201.1380 &   0.74( 45)-0.21( 71)+0.27( 46)+0.49( 30)+0.26( 53)  	  \\
   46 & 2s$^2$2p$^4$($^3$P)3d	&  $^2$F$  _{5/2}$  &201.5632 & 201.4998 &   0.32( 20)+0.60( 71)+0.72( 46)				  \\
   50 & 2s$^2$2p$^4$($^1$D)3d	&  $^2$S$  _{1/2}$  &201.9641 & 202.2051 &  -0.51( 21)-0.39( 59)+0.76( 50)				  \\
   51 & 2s$^2$2p$^4$($^1$D)3d	&  $^2$D$  _{5/2}$  &202.3651 & 202.3860 &  -0.26( 20)+0.33( 71)-0.29( 53)-0.82( 51)			  \\
   52 & 2s$^2$2p$^4$($^1$D)3d	&  $^2$P$  _{3/2}$  &202.2375 & 202.4757 &  -0.27( 43)-0.21( 19)+0.41( 28)+0.31( 47)-0.31( 58)+0.71( 52)  \\
   53 & 2s$^2$2p$^4$($^1$D)3d	&  $^2$F$  _{5/2}$  &202.2375 & 202.5446 &   0.28( 20)+0.51( 30)-0.76( 53)+0.26( 51)			  \\
   58 & 2s$^2$2p$^4$($^1$D)3d	&  $^2$D$  _{3/2}$  &203.4039 & 203.5764 &   0.25( 43)-0.58( 72)-0.71( 58)-0.23( 52)			  \\
   59 & 2s$^2$2p$^4$($^1$D)3d	&  $^2$P$  _{1/2}$  &203.7684 & 203.8744 &  -0.60( 27)-0.70( 59)-0.38( 50)				  \\
   72 & 2s$^2$2p$^4$($^3$P)3d	&  $^2$D$  _{3/2}$  &210.8945 & 211.0131 &  -0.47( 43)-0.26( 19)-0.39( 72)-0.33( 47)+0.64( 31)  	  \\
\\  \hline            								                	 
\end{longtable}   								   					       
			      							   					       
%\vspace*{0.5 cm}													       

%}													       
														       
\begin{flushleft}													       
{\small
NIST:  {\tt http://www.nist.gov/pml/data/asd.cfm} \\
GRASP: Present results with the {\sc grasp} code with 501 level calculations  \\	       
}															       
\end{flushleft} 

%\end{document}
\clearpage
\newpage

\renewcommand{\baselinestretch}{1.0}
\footnotesize
\begin{longtable}{@{\extracolsep\fill}rrrrrrrrrrr@{}}
\caption{Comparison of energy levels of ions with 37 $\le$ Z $\le$ 53 (in Ryd).}
Z  & \multicolumn{2}{c}{Experimental}    & \multicolumn{2}{c}{GRASPa}  & \multicolumn{2}{c}{GRASPb}       &   \multicolumn{2}{c}{FAC}     \\   \hline
Level & 2s$^2$2p$^5$~$^2$P$^o_{1/2}$ & 2s2p$^6$~$^2$S$_{1/2}$ & 2s$^2$2p$^5$~$^2$P$^o_{1/2}$ & 2s2p$^6$~$^2$S$_{1/2}$  & 2s$^2$2p$^5$~$^2$P$^o_{1/2}$ & 2s2p$^6$~$^2$S$_{1/2}$  & 2s$^2$2p$^5$~$^2$P$^o_{1/2}$ & 2s2p$^6$~$^2$S$_{1/2}$  \\ \\
 \hline\\
\endfirsthead\\
\caption[]{(continued)}
Z  & \multicolumn{2}{c}{Experimental}    & \multicolumn{2}{c}{GRASPa}  & \multicolumn{2}{c}{GRASPb}       &   \multicolumn{2}{c}{FAC}   \\  \hline
Level & 2s$^2$2p$^5$~$^2$P$^o_{1/2}$ & 2s2p$^6$~$^2$S$_{1/2}$ & 2s$^2$2p$^5$~$^2$P$^o_{1/2}$ & 2s2p$^6$~$^2$S$_{1/2}$  & 2s$^2$2p$^5$~$^2$P$^o_{1/2}$ & 2s2p$^6$~$^2$S$_{1/2}$  & 2s$^2$2p$^5$~$^2$P$^o_{1/2}$ & 2s2p$^6$~$^2$S$_{1/2}$  \\  \\
\hline\\
\endhead
37 &  4.5962 & 18.3250  &  4.5967 & 18.3236 &   4.5872 & 18.4716 &  4.5881 &  18.4550  \\
38 &  5.1793 & 19.3763  &  5.1765 & 19.3736 &   5.1663 & 19.5230 &  5.1676 &  19.5061  \\      
39 &  5.8072 & 20.4798  &  5.8106 & 20.4818 &   5.7998 & 20.6264 &  5.8014 &  20.6156  \\
40 &  6.5001 & 21.6581  &  6.5028 & 21.6520 &   6.4911 & 21.8044 &  6.4932 &  21.7870  \\
41 &  7.2500 & 22.8765  &  7.2565 & 22.8879 &   7.2441 & 23.0419 &  7.2466 &  23.0244  \\
42 &  8.0756 & 24.1969  &  8.0756 & 24.1035 &   8.0624 & 24.3494 &  8.0655 &  24.3315  \\
43 &	     &          &  8.9641 & 25.5730 &   8.9500 & 25.7308 &  8.9538 &  25.7127  \\
44 &	     &          &  9.9262 & 27.0306 &   9.9112 & 27.1906 &  9.9157 &  27.1721  \\
45 & 10.9753 & 28.5874  & 10.9661 & 28.5709 &  10.9520 & 28.7331 & 10.9555 &  28.7143  \\
46 &	     &          & 12.0883 & 30.1983 &  12.0715 & 30.3631 & 12.0777 &  30.3438  \\
47 &	     & 31.9272  & 13.2975 & 31.9179 &  13.2797 & 32.0854 & 13.2869 &  32.0656  \\
48 &	     & 33.7734  & 14.5985 & 33.7344 &  14.5797 & 33.9050 & 14.5880 &  33.8845  \\
49 &	     &          & 15.9964 & 35.6533 &  15.9764 & 35.8272 & 15.9860 &  35.8059  \\
50 &	     & 37.7137  & 17.4963 & 37.6799 &  17.4753 & 37.8574 & 17.4862 &  37.8351  \\
51 &	     &          & 19.1037 & 39.8195 &  19.0815 & 40.0012 & 19.0939 &  39.9778  \\
52 &	     &          & 20.8243 & 42.0784 &  20.8009 & 42.2645 & 20.8149 &  42.2395  \\
53 &	     &          & 22.6639 & 44.4631 &  22.6393 & 44.6534 & 22.6550 &  44.6272  \\
  \\ \hline  											       								   					      
\end{longtable}

%\vspace*{0.5 cm}                                                                               
                               
\begin{flushleft}                                                                               
                               
{\small
%NIST: {\tt http://www.nist.gov/pml/data/asd.cfm} \\
Experimental: see  J\"{o}nsson  et al. \cite{jag} \\
GRASPa: earlier calculations of J\"{o}nsson et al. \cite{jag} with  the {\sc grasp} code \\
GRASPb: present calculations with  the {\sc grasp} code for 501 levels \\
FAC: present calculations with  the {\sc fac} code for 38~089 levels \\                                                                               
}                                                                                               
                               
\end{flushleft} 
%}                        
%\end{document}

\clearpage
\newpage
\renewcommand{\baselinestretch}{1.0}
\footnotesize
\begin{longtable}{@{\extracolsep\fill}rllrrrrr@{}}
\caption{ Comparison of A-values (s$^{-1}$) for transitions among the lowest three levels of ions with 37 $\le$ Z $\le$ 53. The first entry is from the present calculations with {\sc grasp} and the second is from J\"{o}nsson  et al. \cite{jag}.  $a{\pm}b \equiv a{\times}$10$^{{\pm}b}$.}
Z  & 1--3 (E1)& 2--3 (E1)& 1--2 (M1)& 1--2 (E2) \\  \\
 \hline\\
\endfirsthead\\
\caption[]{(continued)}
Z  & 1--3 (E1)& 2--3 (E1)& 1--2 (M1)& 1--2 (E2)  \\  \\
\hline\\
\endhead 
37 & 2.521+11 & 5.171+10 & 2.274+06 & 1.054+03  \\
37 & 2.419+11 & 4.911+10 & 2.289+06 & 1.051+03  \\
38 & 2.806+11 & 5.379+10 & 3.247+06 & 1.687+03  \\ 
38 & 2.698+11 & 5.116+10 & 3.266+06 & 1.681+03  \\
39 & 3.128+11 & 5.591+10 & 4.591+06 & 2.665+03  \\
39 & 3.012+11 & 5.232+10 & 4.617+06 & 2.656+03  \\
40 & 3.492+11 & 5.806+10 & 6.432+06 & 4.159+03  \\
40 & 3.367+11 & 5.534+10 & 6.468+06 & 4.146+03  \\
41 & 3.903+11 & 6.024+10 & 8.933+06 & 6.419+03  \\
41 & 3.678+11 & 5.747+10 & 8.981+06 & 6.399+03  \\
42 & 4.367+11 & 6.246+10 & 1.231+07 & 9.800+03  \\
42 & 4.222+11 & 5.964+10 & 1.237+07 & 9.770+03  \\
43 & 4.894+11 & 6.472+10 & 1.683+07 & 1.481+04  \\
43 & 4.737+11 & 6.185+10 & 1.691+07 & 1.477+04  \\
44 & 5.490+11 & 6.701+10 & 2.283+07 & 2.217+04  \\
44 & 5.320+11 & 6.409+10 & 2.294+07 & 2.210+04  \\
45 & 6.165+11 & 6.935+10 & 3.077+07 & 3.288+04  \\
45 & 5.981+11 & 6.637+10 & 3.091+07 & 3.278+04  \\ 
46 & 6.931+11 & 7.172+10 & 4.119+07 & 4.834+04  \\
46 & 6.731+11 & 6.868+10 & 4.137+07 & 4.819+04  \\
47 & 7.800+11 & 7.415+10 & 5.479+07 & 7.047+04  \\
47 & 7.582+11 & 7.104+10 & 5.503+07 & 7.027+04  \\
48 & 8.786+11 & 7.661+10 & 7.245+07 & 1.019+05  \\
48 & 8.548+11 & 7.343+10 & 7.275+07 & 1.016+05  \\
49 & 9.906+11 & 7.913+10 & 9.524+07 & 1.463+05  \\
49 & 9.645+11 & 7.587+10 & 9.563+07 & 1.459+05  \\
50 & 1.118+12 & 8.169+10 & 1.245+08 & 2.086+05  \\
50 & 1.089+12 & 7.836+10 & 1.250+08 & 2.080+05  \\
51 & 1.262+12 & 8.430+10 & 1.620+08 & 2.952+05  \\
51 & 1.231+12 & 8.088+10 & 1.626+08 & 2.944+05  \\
52 & 1.426+12 & 8.697+10 & 2.096+08 & 4.150+05  \\
52 & 1.391+12 & 8.346+10 & 2.104+08 & 4.139+05  \\
53 & 1.612+12 & 8.969+10 & 2.700+08 & 5.799+05  \\
53 & 1.574+12 & 8.609+10 & 2.709+08 & 5.783+05  \\
\\  \hline            								                	 
\end{longtable}   								   					       
			      							   					       
%\vspace*{0.5 cm}													       

%}													       
														       
\begin{flushleft}													       
{\small
		      		      	       
}															       
\end{flushleft} 

\clearpage
\newpage
\renewcommand{\baselinestretch}{1.0}
\footnotesize
\begin{longtable}{@{\extracolsep\fill}rrrrrrrrrrrrr@{}}
\caption{ Comparison of f-values (dimensionless) for transitions from the ground (2s$^2$2p$^5$~$^2$P$^o_{3/2}$) to higher excited levels of Mo~XXXIV, Pd~XXXVIII and Sn~XLII.  See Tables 6, 10 and 14 for definitions of J levels. $a{\pm}b \equiv a{\times}$10$^{{\pm}b}$.}
     &   \multicolumn{3}{c}{Mo~XXXIV} &   \multicolumn{3}{c}{Pd~XXXVIII} &   \multicolumn{3}{c}{Sn~XLII}     \\   \hline
      I  &    J  &    GRASP    &   DFS    &    J &  GRASP      &    DFS    &    J &  GRASP     &   DFS     \\
 \hline\\
\endfirsthead\\
\caption[]{(continued)}
     &   \multicolumn{3}{c}{Mo~XXXIV} &   \multicolumn{3}{c}{Pd~XXXVIII} &   \multicolumn{3}{c}{Sn~XLII}   \\    \hline
     I  &    J  &    GRASP    &   DFS    &    J &  GRASP      &    DFS    &    J &  GRASP     &   DFS      \\
\hline\\
\endhead
     1  &    3  &  4.585$-$02 &  0.0492  &    3 &  4.680$-$02 &   0.0499  &    3 & 4.854$-$02 & 0.0514  \\
     1  &    4  &  1.300$-$02 &  0.0121  &    4 &  1.392$-$02 &   0.0130  &    4 & 1.462$-$02 & 0.0137  \\
     1  &    5  &  6.581$-$02 &  0.0611  &    5 &  6.506$-$02 &   0.0608  &    5 & 6.474$-$02 & 0.0609  \\
     1  &    6  &  1.494$-$02 &  0.0147  &    6 &  1.516$-$02 &   0.0150  &    6 & 1.532$-$02 & 0.0152  \\
     1  &   13  &  1.532$-$02 &  0.0143  &   15 &  1.681$-$02 &   0.0158  &   15 & 1.063$-$02 & 0.0079  \\
     1  &   15  &  1.175$-$02 &  0.0110  &   16 &  1.220$-$02 &   0.0115  &   16 & 1.221$-$03 & 0.0158  \\ 
     1  &   17  &  5.087$-$02 &  0.0474  &   17 &  6.029$-$02 &   0.0570  &   17 & 1.489$-$02 & 0.0011  \\
     1  &   18  &  7.820$-$03 &  0.0074  &   18 &  1.072$-$02 &   0.0102  &   18 & 1.595$-$02 & 0.0176  \\ 
     1  &   19  &  8.551$-$04 &  0.0009  &   19 &  1.744$-$03 &   0.0018  &   20 & 2.593$-$02 & 0.0346  \\
     1  &   20  &  3.169$-$04 &  0.0002  &   20 &  1.967$-$03 &   0.0017  &   23 & 1.242$-$01 & 0.1191  \\
     1  &   21  &  8.687$-$03 &  0.0087  &   21 &  1.418$-$02 &   0.0140  &   24 & 3.756$-$02 & 0.0241  \\ 
     1  &   27  &  1.114$-$01 &  0.1070  &   25 &  1.144$-$01 &   0.1100  &   25 & 1.064$-$01 & 0.1021  \\ 
     1  &   28  &  1.901$-$01 &  0.1811  &   26 &  1.619$-$01 &   0.1535  &   26 & 9.281$-$02 & 0.0893  \\ 
     1  &   30  &  4.707$-$01 &  0.4578  &   27 &  4.934$-$01 &   0.4816  &   27 & 4.625$-$01 & 0.4532  \\
     1  &   31  &  2.358$-$01 &  0.2379  &   29 &  2.923$-$01 &   0.2943  &   28 & 3.501$-$01 & 0.3500  \\
     1  &   35  &  4.130$-$01 &  0.4304  &   31 &  4.497$-$01 &   0.4673  &   29 & 4.643$-$01 & 0.4813  \\
     1  &   41  &  2.699$-$04 &  0.0003  &   41 &  7.399$-$04 &   0.0008  &   41 & 8.508$-$05 & 0.0001  \\   
     1  &   42  &  2.548$-$04 &  0.0002  &   42 &  2.752$-$04 &   0.0002  &   42 & 7.897$-$03 & 0.0077  \\
     1  &   43  &  8.816$-$03 &  0.0085  &   43 &  8.227$-$03 &   0.0080  &   43 & 1.905$-$01 & 0.1866  \\
     1  &   45  &  2.125$-$01 &  0.2072  &   44 &  1.995$-$01 &   0.1951  &   47 & 9.784$-$03 & 0.0012  \\
     1  &   46  &  1.407$-$02 &  0.0141  &   46 &  8.456$-$03 &   0.0083  &   49 & 1.753$-$03 & 0.0094  \\
     1  &   47  &  1.792$-$03 &  0.0015  &   48 &  1.520$-$03 &   0.0012  &   50 & 1.616$-$01 & 0.1620  \\
     1  &   50  &  2.161$-$01 &  0.2020  &   50 &  2.151$-$01 &   0.2029  &   51 & 5.637$-$01 & 0.5749  \\ 
     1  &   51  &  5.732$-$02 &  0.0511  &   52 &  4.455$-$01 &   0.5083  &   52 & 4.442$-$01 & 0.4406  \\
     1  &   52  &  4.496$-$01 &  0.4458  &   53 &  5.196$-$01 &   0.4418  &   54 & 5.633$-$02 & 0.0452  \\ 
     1  &   53  &  5.757$-$01 &  0.6003  &   54 &  7.528$-$02 &   0.1007  &   55 & 3.149$-$03 & 0.0033  \\
     1  &   58  &  7.751$-$02 &  0.0794  &   57 &  5.603$-$02 &   0.0575  &   57 & 3.939$-$02 & 0.0406  \\
     1  &   59  &  3.292$-$02 &  0.0368  &   58 &  2.554$-$02 &   0.0287  &   60 & 1.767$-$02 & 0.0203  \\ 
     1  &   63  &  7.920$-$03 &  0.0078  &   62 &  8.892$-$03 &   0.0088  &   61 & 8.400$-$03 & 0.0084  \\
     1  &   64  &  5.384$-$02 &  0.0539  &   63 &  5.365$-$02 &   0.0537  &   62 & 4.960$-$02 & 0.0497  \\
     1  &   66  &  2.971$-$02 &  0.0276  &   65 &  9.826$-$05 &   0.0000  &   64 & 2.454$-$02 & 0.0252  \\
     1  &   67  &  7.246$-$02 &  0.0732  &   66 &  2.333$-$02 &   0.0240  &   65 & 1.200$-$02 & 0.0127  \\
     1  &   68  &  1.748$-$02 &  0.0182  &   68 &  1.104$-$01 &   0.1098  &   67 & 9.709$-$02 & 0.0738  \\ 
     1  &   69  &  7.904$-$02 &  0.0805  &   69 &  7.317$-$02 &   0.0738  &   68 & 7.327$-$02 & 0.0996  \\
     1  &   70  &  5.894$-$02 &  0.0598  &   70 &  5.436$-$02 &   0.0551  &   69 & 5.310$-$02 & 0.0538  \\ 
     1  &   71  &  4.054$-$10 &  0.0000  &   71 &  8.086$-$02 &   0.0816  &   71 & 8.477$-$02 & 0.0852  \\
     1  &   72  &  2.619$-$03 &  0.0024  &   72 &  3.054$-$02 &   0.0329  &   72 & 3.161$-$02 & 0.0339  \\
     1  &   74  &  8.011$-$02 &  0.0810  &   74 &  2.293$-$03 &   0.0028  &   73 & 2.022$-$03 & 0.0030  \\
     1  &   75  &  2.788$-$02 &  0.0301  &   75 &  6.354$-$07 &   0.0000  &   83 & 1.027$-$05 & 0.0000  \\
     1  &   76  &  2.602$-$03 &  0.0037  &   76 &  2.842$-$03 &   0.0033  &   84 & 6.129$-$04 & 0.0007  \\ 
     1  &   86  &  4.684$-$04 &  0.0007  &   90 &  1.924$-$04 &   0.0003  &   94 & 9.104$-$05 & 0.0001  \\
     1  &   92  &  6.058$-$04 &  0.0008  &   95 &  9.307$-$04 &   0.0012  &   95 & 4.662$-$04 & 0.0006  \\
     1  &   93  &  1.236$-$03 &  0.0015  &   96 &  3.450$-$07 &   0.0000  &   96 & 4.278$-$05 & 0.0000  \\
     1  &   97  &  5.227$-$06 &  0.0001  &   97 &  3.321$-$05 &   0.0001  &   97 & 2.698$-$06 & 0.0001  \\ 
     1  &   98  &  9.719$-$05 &  0.0003  &   98 &  3.715$-$05 &   0.0001  &   98 & 1.474$-$05 & 0.0001  \\
     1  &   99  &  6.515$-$03 &  0.0066  &   99 &  3.580$-$03 &   0.0036  &   99 & 1.995$-$03 & 0.0020  \\
     1  &  100  &  1.665$-$03 &  0.0024  &  100 &  8.723$-$04 &   0.0013  &  100 & 4.602$-$04 & 0.0007  \\
     1  &  109  &  1.467$-$04 &  0.0003  &  109 &  1.220$-$04 &   0.0002  &  109 & 9.614$-$05 & 0.0002  \\ 
     1  &  112  &  7.993$-$05 &  0.0001  &  112 &  6.248$-$05 &   0.0001  &  112 & 4.736$-$05 & 0.0001  \\
     1  &  113  &  7.549$-$04 &  0.0013  &  113 &  5.908$-$04 &   0.0010  &  113 & 4.508$-$04 & 0.0008  \\
 \\ \hline  											       								   					      
\end{longtable}

%\vspace*{0.5 cm}                                                                               
                               
\begin{flushleft}                                                                               
                               
{\small
GRASP: present calculations with the {\sc grasp} code \\
DFS: earlier calculations of Sampson et al. \cite{sam} with the Dirac-Fock-Slater code \\                                                                               
}                                                                                               
                               
\end{flushleft} 
%}                        
\end{document}